\documentclass[%
 reprint, %onecolumn,
 amsmath,amssymb,
 aps,
]{revtex4-2}
\usepackage{}

\usepackage{graphicx}% Include figure files
\usepackage{dcolumn}% Align table columns on decimal point
\usepackage{bm}% bold math
\usepackage{epstopdf}
\usepackage{float}
\usepackage{xcolor, colortbl}

\newcommand{\beq}{\vspace{0.5em}\begin{equation}}
\newcommand{\eeq}{\end{align}\vspace{0.5em}}
\newcommand{\beqn}{\vspace{0.5em}\begin{eqnarray}}
\newcommand{\eeqn}{\end{eqnarray}\par\vspace{0.5em}\noindent}

\newcommand{\bsub}{\begin{subequations}}
\newcommand{\esub}{\end{subequations}}

\begin{document}

\preprint{APS/123-QED}

\title{Coupling of shape and pairing vibrations in a collective Hamiltonian based on nuclear energy density functionals}% Force line breaks with \\

\author{J. Xiang}
\affiliation{School of Physics and Electronic, Qiannan Normal University for Nationalities, Duyun, 558000, China}
\affiliation{School of Physical Science and Technology, Southwest University, Chongqing 400715, China}
\author{Z. P. Li}
\email{zpliphy@swu.edu.cn}
\affiliation{School of Physical Science and Technology, Southwest University, Chongqing 400715, China}

\author{T. Nik\v si\' c}
\affiliation{Physics Department, Faculty of Science, University of Zagreb, Croatia}
\author{D. Vretenar}
\affiliation{Physics Department, Faculty of Science, University of Zagreb, Croatia}

\author{W. H. Long}
\affiliation{School of Nuclear Science and Technology, Lanzhou University, Lanzhou 730000, China}

\date{\today}% It is always \today, today,

\begin{abstract}
The quadrupole collective Hamiltonian, based on relativistic energy density functionals, is extended to include a pairing collective coordinate. In addition to quadrupole shape vibrations and rotations, the model describes pairing vibrations and the coupling between shape and pairing degrees of freedom. The parameters of the collective Hamiltonian are determined by constrained self-consistent relativistic mean-field plus Bardeen-Cooper-Schrieffer (RMF+BCS) calculations in the space of intrinsic shape and pairing deformations. The effect of coupling between shape and pairing degrees of freedom is analyzed in a study of low-energy spectra and transition rates of four axially symmetric $N=92$ rare-earth isotones. When compared to results obtained with the standard quadrupole collective Hamiltonian, the inclusion of dynamical pairing increases the moment of inertia, lowers the energies of excited $0^+$ states and reduces the E0-transition strengths, in better agreement with data.
\end{abstract}

\maketitle

%\tableofcontents

\section{\label{sec:I}Introduction}

Atomic nuclei are finite-size, strongly correlated quantum many-body systems, and their complex spectra exhibit a variety of excitation modes determined by collective and single-particle degrees of freedom \cite{Bohr_1981,WambachNPA1982}. For low-energy excitation spectra, in particular, the coupling between shape degrees of freedom and two-quasiparticle excitations play an important role \cite{SiegalNPA1972,MauroPRC1987,NuriaPLB2011,NurialopezPRC2013}. The occurrence of pairing vibrations in nuclei was suggested by Bohr and Mottelson~\cite{Bohr1964}, and this mode influences many physical quantities in addition to low-energy spectra, such as nuclear matrix elements for neutrinoless double-beta decay \cite{NurialopezPRL2013} and spontaneous fission half-lifes \cite{SadhukhanPRC2014,ZhaoPRC2016}.

The pairing interaction between nucleons produces correlations that enhance the amplitude of two-nucleon transfer. Thus, two-nucleon transfer reactions provide a tool to identify pairing excitations, and a number of pairing vibrational states have been observed in heavier nuclei, e.g. the proton pairing vibrational states with excitation energy: $5.24$ MeV in $^{208}$Pb \cite{IGO1971AnP,Anderson1977PRL}, $4.1$ MeV in $^{206}$Pb \cite{Anderson1977PRL}, $1.690$ MeV in $^{124}$Xe,  $1.761$ MeV in
$^{126}$Xe \cite{AlfordNPA1979,RadichPRC2015}.

A variety of theoretical methods have been used to describe pairing vibrations: the pairing Hamiltonian~\cite{BesNP1966,BrogliaNPA1967,BrogliaNPA1968}, the collective Hamiltonian~\cite{BESNPA1970,DusselNPA1971,DusselNPA1972}, the time-dependent Hartree-Fock-Bogoliubov theory~\cite{AvezPRC2008,RipkaNPA1969}, the shell model~ \cite{HeuslerPRC2015}, the quasiparticle random phase approximation \cite{HirotakaPRC2011,KhanPRC2009,SharmaPRC1994,Dussel1988PRC,BesNP1966,BrogliaNPA1968b,GoswamiPRC1970}, the pair addition and pair removal phonon model~ \cite{JolosPRC2015}, and the generator coordinate method (GCM)~\cite{RipkaNPA1969,SiegalNPA1972,FaesslerNPA1973,GozdzNPA1985,PomorskiAPPB1987,SiejaEPJA2004,ProchniakIJMPE2007}. In general, however, these methods have not explicitly considered the coupling between shape and pairing vibrations.

In Ref. \cite{PomorskiIJMPE2007} the GCM with the Gaussian overlap approximation was extended to include both pairing and shape vibrations, and their coupling. It was shown that the energy of the lowest excited $0^+$ state of $^{120,124-130}$Xe is considerably reduced as a result of the inclusion of pairing vibrations. Vaquero {\it et al.} have explicitly considered the collective pairing degree of freedom using a finite range force in the
framework of the symmetry conserving configuration mixing (SCCM) approach \cite{NuriaPLB2011}. They have also extend the analysis to other observables like transition probabilities and separation energies \cite{NurialopezPRC2013}.
The method has been used to compute nuclear matrix elements of neutrinoless double beta decay\cite {NurialopezPRL2013}. An increase of 10\%--40\% in the nuclear matrix elements with respect to the ones calculated without the inclusion of pairing fluctuations has been obtained, reducing the predicted half-lives of these isotopes. However, the numerical implementation of the model is very involved, and applications to medium-heavy and heavy nuclei are computationally excessive.

The low-energy structure of medium-heavy and heavy nuclei is best described in the framework of nuclear energy density functionals (EDFs)
\cite{BenderRMP2003,VretenarPhyRep2005,MengPPNP2006,StonePPNP2007,Mengbook2016,ZhaoPLB2011,ZhaoPRC2012,ZhaoPRL2011,NiuPLB2013,NiuPRC2009,NiuPRC2013}. The basic implementation of this framework, in which an EDF is constructed as a functional of one-body nucleon density, is in terms of self-consistent mean-field (SCMF) models. To calculate excitation spectra and electromagnetic transition rates  the SCMF method must be extended to include collective correlations that arise from symmetry restoration and fluctuations around the mean- field minima. A particularly convenient approach is the collective Hamiltonian with parameters and the potential determined by SCMF calculations. Over the last decade a five-dimensional collective Hamiltonian (5DCH) model for quadrupole vibrational and rotational degrees of freedom, based on relativistic density functionals, has been developed and applied in a number of studies of structure phenomena related to shape coexistence and shape transitions
\cite{NiksicPRC2009,LiPRC2009a,LiPRC2009b,LiPRC2010b,LiPRC2011,NiksicPPNP2011,LiJPG2016,
LuPRC2015,XiangPRC2016,XiangPRC2018}. Another development is the exploration of  quadrupole and octupole vibrations, rotations, and their coupling using the quadrupole-octupole collective Hamiltonian (QOCH). With parameters determined by relativistic nuclear energy density functionals, the QOCH has successfully been applied to
systematic studies of quadrupole and octupole states in even-even medium-heavy and heavy
nuclei \cite{LiPLB2013,LiJPG2016,XiaPRC2017,XuCPC2017,TaoPRC2017}.

In various implementations of our collective Hamiltonian model only shape and rotational degrees of freedom have been considered as collective coordinates. Pairing correlations have been taken into account on the SCMF level, either in the relativistic mean-field RMF plus BCS approximation, or the relativistic Hartree-Bogoliubov framework. In the present study we develop the quadrupole-and-pairing collective Hamiltonian (QPCH) that, in addition to quadrupole shape vibrations and rotations, includes pairing vibrations and explicitly couples shape and pairing degrees of freedom.

The paper is organized as follows. Section \ref{Sec:II} outlines the theoretical framework, in particular the calculation of mass parameters and moments of inertia entering the QPCH, and the method of solution of the QPCH eigenvalue problem. In Sec. \ref{Sec:III} we analyze the effect of coupling between shape and pairing degrees of freedom on the low-energy spectra and transition rates of four axially symmetric $N=92$ rare-earth isotones. Section \ref{Sec:IIII} presents a summary and an outlook for future studies.

%%%%%%%%%%%%%%%%%%%%%%%%%%%%%%%%%%%%%%%%%%%%%%%%%%%%%

\section{\label{Sec:II}Theoretical Framework}

 %
%\subsection{Collective Hamiltonian for the coupling of shape and pairing vibrations}
For a description of pairing vibrations the monopole pairing operator can be defined in the following form \cite{PomorskiIJMPE2007,BESNPA1970}
\begin{align}
\hat{A} = \frac{1}{2}\sum_{k>0}(e^{-2i\phi}c_kc_{\bar{k}}+e^{2i\phi}c^\dagger_{\bar{k}}c^\dagger_{k}),
\end{align}
where $\phi$ is the gauge angle.
The expectation value of this operator in the BCS-like state
\begin{align}
\mid\alpha\phi\rangle=e^{iN\phi}\prod\limits_{k>0}\left(u_k+v_ke^{-2i\phi} c^\dagger_kc^\dagger_{\bar{k}}\right)\mid0\rangle,
\end{align}
determines the pair condensate.
In this work we do not consider the so called ``pairing rotations'', that is, quasi-rotational bands that correspond to ground states of neighboring even-even nuclei. Therefore, the gauge angle can be chosen $\phi=0$, and the pairing operator  reduces to
\begin{align}
\label{Eq:P-operator}
\hat{P} = \frac{1}{2}\sum_{k>0}(c_kc_{\bar{k}}+c^\dagger_{\bar{k}}c^\dagger_{k}).
\end{align}
The mean value of this operator:
\begin{align}
\alpha_\tau = \langle \alpha (\phi=0) \mid\hat{P}\mid \alpha (\phi=0) \rangle_{\tau}=\sum_{k>0}u^{\tau}_kv^{\tau}_k,
\end{align}
with $\tau$ denoting neutron or proton states, defines the intrinsic pairing deformation $\alpha$ related to the pairing gap parameter $\Delta$
\begin{align}
\alpha=\sum\limits_{\tau=n,p}\sum_{k>0}u^{\tau}_kv^{\tau}_k.
\end{align}
In the following, $\alpha$ will be considered as the pairing collective coordinate.
Nuclear excitations characterised by axially symmetric quadrupole shape vibrational and rotational collective motion, and  coupled with pairing vibrations, can be described by constructing a collective Hamiltonian defined by the quadrupole shape deformation parameter $\beta$, the Euler angle $\Omega$, and pairing deformation $\alpha$ as collective coordinates (denoted as QPCH). The quantized collective Hamiltonian takes the general form
\begin{align}
   \label{eq:QPCH}
{\hat H}_{\rm coll} = -\frac{\hbar^2}{2\sqrt{g{\cal I}}}\sum\limits_{i,j}\frac{\partial}{\partial q_i}\sqrt{g{\cal I}}(B^{-1})_{ij}\frac{\partial}{\partial q_j}
+\frac{\hat{J}^2}{2{\cal I}}+{V}_{\rm coll}(q),
\end{align}
where the collective mass tensor reads
\begin{align}
B=\left(
\begin{array}{cc}
B_{\beta\beta}       & B_{\beta\alpha} \\
B_{\alpha\beta}   & B_{\alpha\alpha}\\
\end{array}
\right),
\end{align}
and $g={\rm det} B$. ${\cal I}$ and ${V}_{\rm coll}$ are the moment of inertia and collective potential, respectively. The corresponding volume element in the collective space reads
\begin{align}
\int d\tau_{\rm coll} = \int\sqrt{g{\cal I}}d\beta d\alpha d\Omega.
\end{align}

To solve the eigenvalue problem for the collective Hamiltonian of Eq.~$(\ref{eq:QPCH})$, the eigenfunctions are expanded in terms of a complete set of basis functions. For each value of the angular momentum $I$ the basis is constructed as:
\begin{align}
\label{eq:basis}
|n_1n_2IMK\rangle=(g{\cal I})^{-1/4}\phi_{n_1}(\beta)\phi_{n_2}(\alpha)|IMK\rangle,
\end{align}
where $\phi_{n_i}$ denotes the one-dimensional harmonic oscillator eigenstate for the corresponding collective coordinate. The present study is restricted by axial symmetry, and thus the projection of angular momentum $K=0$. The collective wave function can finally be written as
\begin{align}\label{collwavefunc}
\Psi^{IM}_j(\beta, \alpha, \Omega)=\psi^{I}_j(\beta, \alpha)|IM0\rangle.
\end{align}

The reduced transition rates can be computed using the expression
\begin{align}
B(E\lambda, I_i\to I_f)&=
\langle I_i0\lambda0|I_f0\rangle^2 \nonumber \\ &\times  \left|\int d\beta d\alpha \sqrt{g{\cal I}} \psi_i
         \mathcal{M}_{E\lambda}(\beta, \alpha) \psi^*_{f}\right|^2,
\end{align}
where $\mathcal{M}_{E\lambda}(\beta, \alpha)$ denotes the electric moment of order $\lambda$. In microscopic models it is calculated as $\langle\Phi(\beta, \alpha)|\hat{\mathcal{M}}(E\lambda)|\Phi(\beta, \alpha)\rangle$, where $\Phi(\beta, \alpha)$ is the nuclear wave function.

For comparison we will also consider a pairing collective Hamiltonian (PCH) that describes one-dimensional pairing vibrational motion in $\alpha$:
\begin{equation}\label{eq:PCH}
{\hat H}_{\rm coll}=-\frac{\hbar^2}{2}\frac{1}{\sqrt{B_{\alpha\alpha}}}\frac{\partial}{\partial \alpha}\frac{1}{\sqrt{B_{\alpha\alpha}}}\frac{\partial}{\partial\alpha} + V_{\rm coll}(\alpha)\;.
\end{equation}

The entire dynamics of the collective Hamiltonian Eq.~(\ref{eq:QPCH}) is governed by the five functions of the intrinsic quadrupole deformation $\beta$ and pairing deformation $\alpha$: the collective potential, the three mass parameters $B_{\beta\beta}$, $B_{\alpha\alpha}$, $B_{\beta\alpha}$, and the moment of inertia $\mathcal{I}$. These functions are determined by the choice of a particular microscopic nuclear energy density functional and pairing interaction. In the present study the energy density functional PC-PK1 \cite{ZhaoPRC2010} determines the effective interaction in the particle-hole channel, and the Bardeen-Cooper-Schrieffer (BCS) approximation with a separable pairing force is employed in the particle-particle channel
\cite{TianPLB2009,NIksicPRC2010}. The framework of the relativistic mean-field model plus BCS (RMF+BCS) with a separable pairing force is described in detail in Ref. \cite{Xiang2012NPA}.

The map of the collective energy surface as a function of $\beta$ and $\alpha$ is obtained by imposing constraints on the mass quadrupole moment $q$ and pairing deformation $\alpha$, respectively \cite{RingBook,SiejaEPJA2004}. The method of quadratic constraint on mass quadrupole moment uses an unrestricted variation of the function
\begin{align}
\langle H\rangle+\frac{1}{2}C_\beta\left(\langle \hat{Q}  \rangle - q  \right)^2
-\lambda\langle\hat{N}-N\rangle
-\xi_\alpha\langle\hat{P}-\alpha\rangle,
\label{constr}
\end{align}
where $\langle H\rangle$ is the total energy, and  $\langle \hat{Q}\rangle$ denotes the expectation value of the mass quadrupole operator:
\begin{align}
\hat{Q}=2z^2-x^2-y^2 \;.
\end{align}
$q$ is the constrained value of the quadrupole moment, and $C_\beta$ the corresponding stiffness constant \cite{RingBook}. The quadrupole deformation parameter $\beta$ is calculated from: $\beta=\frac{\sqrt{5\pi}}{3AR^2_0}q$, with $R_0=r_0A^{1/3}$ and $r_0=1.2$ fm. $\hat{N}$ is the particle number operator, while $\hat{P}$ is the pairing operator defined in Eq. (\ref{Eq:P-operator}). $\lambda$ and $\xi_\alpha$ are Lagrange multipliers. $N$ and $\alpha$ are the constrained value of particle number and pairing deformation, respectively.

The single-nucleon wave functions, energies and occupation probabilities, generated from constrained self-consistent solutions of the RMF+BCS equations, provide the microscopic input for the parameters of the collective Hamiltonian.

The moments of inertia are calculated according to the Inglis-Belyaev formula:~\cite{InglisPR1956,BeliaevNP1961}
\begin{align}
\label{eq:MOI}
\mathcal{I} = \sum_{i,j}{\frac{\left(u_iv_j-v_iu_j \right)^2}{E_i+E_j}
  | \langle i |\hat{J} | j  \rangle |^2},
\end{align}
where $\hat J$ is the angular momentum along the axis perpendicular to the symmetry axis, and the summation runs over the proton and neutron quasiparticle states. The quasiparticle energies $E_i$, occupation probabilities $v_i$, and single-nucleon wave functions $\psi_i$ are determined by solutions of the constrained RMF+BCS equations.

The cranking approximation \cite{GirodNPA1979,PomorskiIJMPE2007} is used for the mass parameters:
\begin{align}
B_{\beta\beta}&=\hbar^2\left[\mathcal {M}^{-1}_{(1)}\mathcal{M}_{(3)}\mathcal{M}^{-1}_{(1)}\right],\\
B_{\alpha\alpha}&=\hbar^2\sum\limits_{i>0}\frac{(u^2_i-v^2_i)^2}{8E^3_i}
\sigma^{-2},\\
B_{\beta\alpha}&=-\hbar^2\sum_{i>0}\mathcal {M}^{-1}_{(1)}\frac{u_{i}v_{i}\langle i\mid {\hat Q}\mid i\rangle}{2E_i^2}
\frac{\left(u_{i}^2-v_{i}^2\right)}{2E_i}\sigma^{-1}
\end{align}
with
\begin{align}
\label{masspar-M}
\mathcal{M}_{(n)}&=2\sum_{i>0,j>0}
 {\frac{\left\langle i\right|\hat{Q}\left| j\right\rangle
 \left\langle j\right|\hat{Q}\left| i\right\rangle}
 {(E_i+E_j)^n}\left(u_i v_j+ v_i u_j \right)^2},\\
\sigma&=\sum\limits_{i>0}\frac{(u^2_i-v^2_i)^2}{4E_i}.
\end{align}
For the collective potential $V_{\rm coll}$ in Eq.~(\ref{eq:QPCH}) the total constrained RMF+BCS energy is used. Following  the procedure described in Ref. \cite{PilatNPA1993}, the zero-point energy is not considered in the present version of the model.

%%%%%%%%%%%%%%%%%%%%%%%%%%%%%%%%%%%%%%%%%%%%%%%%%%%%%

\section{\label{Sec:III}Results and discussions}

To test the model that couples shape and pairing vibrations we will perform several illustrative calculations of potential energy surfaces, inertia tensors, and the resulting collective excitation spectra of four even-even rare-earth $N=92$ isotones. In the present RMF+BCS calculation the strength of the separable pairing force is enhanced by 6\% compared to the original value determined in \cite{TianPLB2009,NIksicPRC2010}, namely we use $G=-771.68$ MeV fm$^3$. It has  been shown
that by increasing the pairing strength of the order of few percents, the RMF+BCS model accurately reproduces
results obtained with the full relativistic Hartree-Bogoliubov calculation and the original pairing force \cite{XiangPRC2013}.

%--------------------------------------------------------------
\subsection{The Pairing Collective Hamiltonian (PCH) for $^{156}$Gd}

In Ref.~\cite{SiejaEPJA2004} the generator coordinate method with the Gaussian overlap approximation (GCM+GOA) approach, based on the single-particle Nilsson potential and the $\delta-$pairing interaction, was used to calculate pairing vibrational excitations of $^{148}$Ce. It has been shown that the first excited collective pairing vibrational states for even-even nuclei in the rare-earth region appear at $\approx 2.5$ MeV for protons and $\approx 4.5$ MeV for neutrons. It was also noted that pairing vibrations are strongly coupled to shape degrees of freedom. Here we perform a similar calculation using an EDF-based collective Hamiltonian. In the first step only one-dimensional pairing vibrations are considered in the model.
To this end the equilibrium minimum for $^{156}$Gd is determined using the self-consistent RMF+BCS method: $\beta_{\rm min}=0.325$ and $\Delta_p (\Delta_n)=0.799\ (0.665)$ MeV. The constrained calculations for $\alpha$ in the interval $2\leqslant\alpha\leqslant 50$, with a step of $2.0$, correspond to the fixed equilibrium value $\beta=0.325$.

\begin{figure}[ht]
\centering{\includegraphics[width=0.45\textwidth]{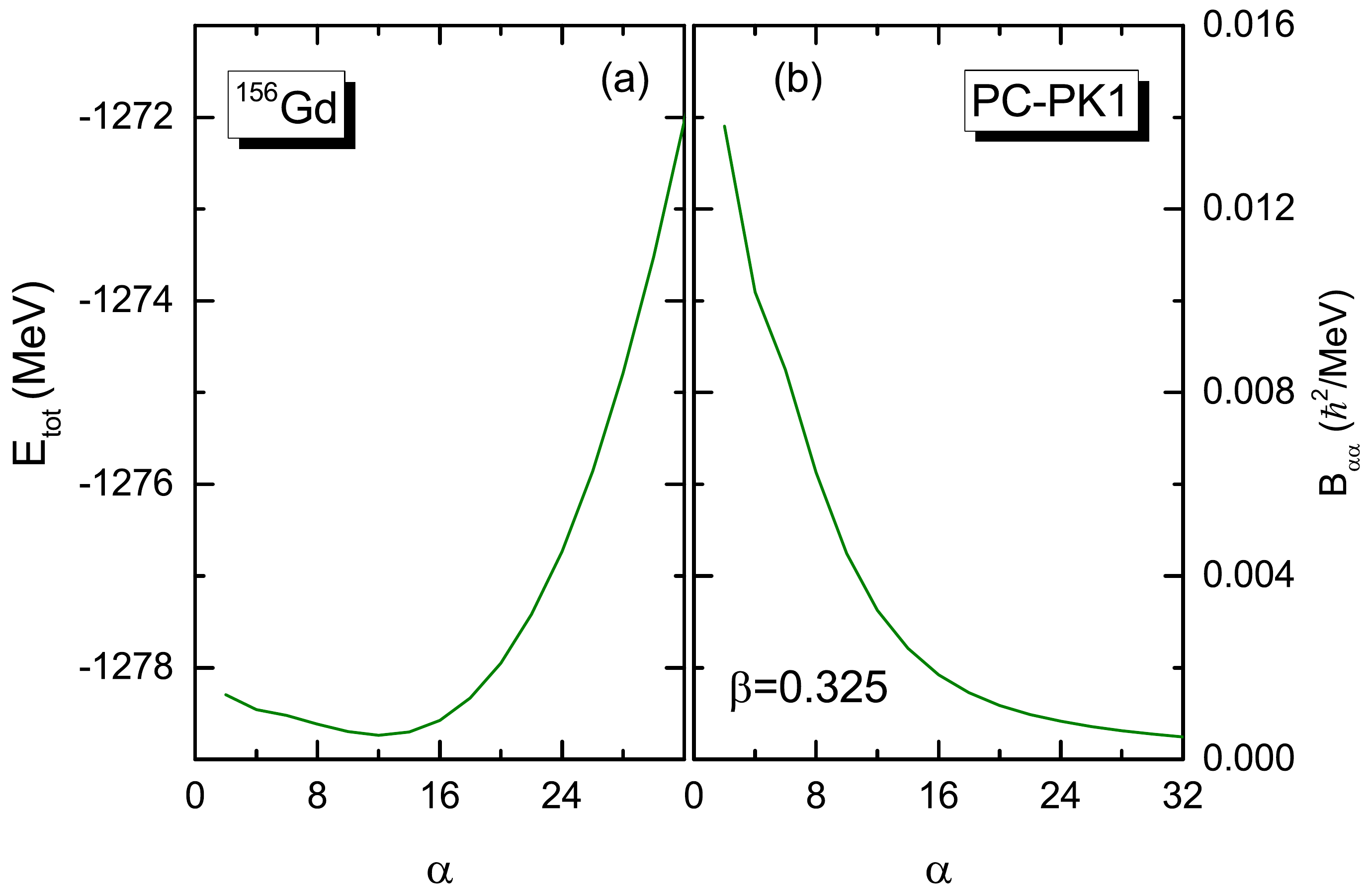}}
\caption{\label{Colpar-Gd156}(Color online) The RMF+BCS binding energy (a) and collective mass $B_{\alpha\alpha}$ (b) of $^{156}$Gd, as functions of the intrinsic pairing deformation $\alpha$.}
\end{figure}

Figure \ref{Colpar-Gd156} displays the RMF+BCS binding energy and collective mass $B_{\alpha\alpha}$ of $^{156}$Gd, as functions of $\alpha$, calculated with the PC-PK1 energy density functional \cite{ZhaoPRC2010} and separable pairing interaction. The deformation energy curve is rather soft for $\alpha \leq 16$ with a shallow minimum at $\alpha\sim12$, and then increases steeply for stronger pairing. The collective mass $B_{\alpha\alpha}$, in contrast, exhibits a steep decrease with $\alpha$. Only for larger values of $\alpha$ ($\alpha > 16$) this decrease becomes more gradual. Such a functional dependence of $B_{\alpha\alpha}$ makes it difficult to obtain converged numerical solutions for the pairing collective Hamiltonian of Eq. (\ref{eq:PCH}). Following the method of Refs. \cite{GozdzNPA1985,SiejaEPJA2004}, we thus use a logarithmic function to transform $\alpha$ into a new coordinate. Details of this transformation are described in Appendix \ref{Sec:ADCH}.

\begin{figure}[ht]
\includegraphics[width=0.45\textwidth]{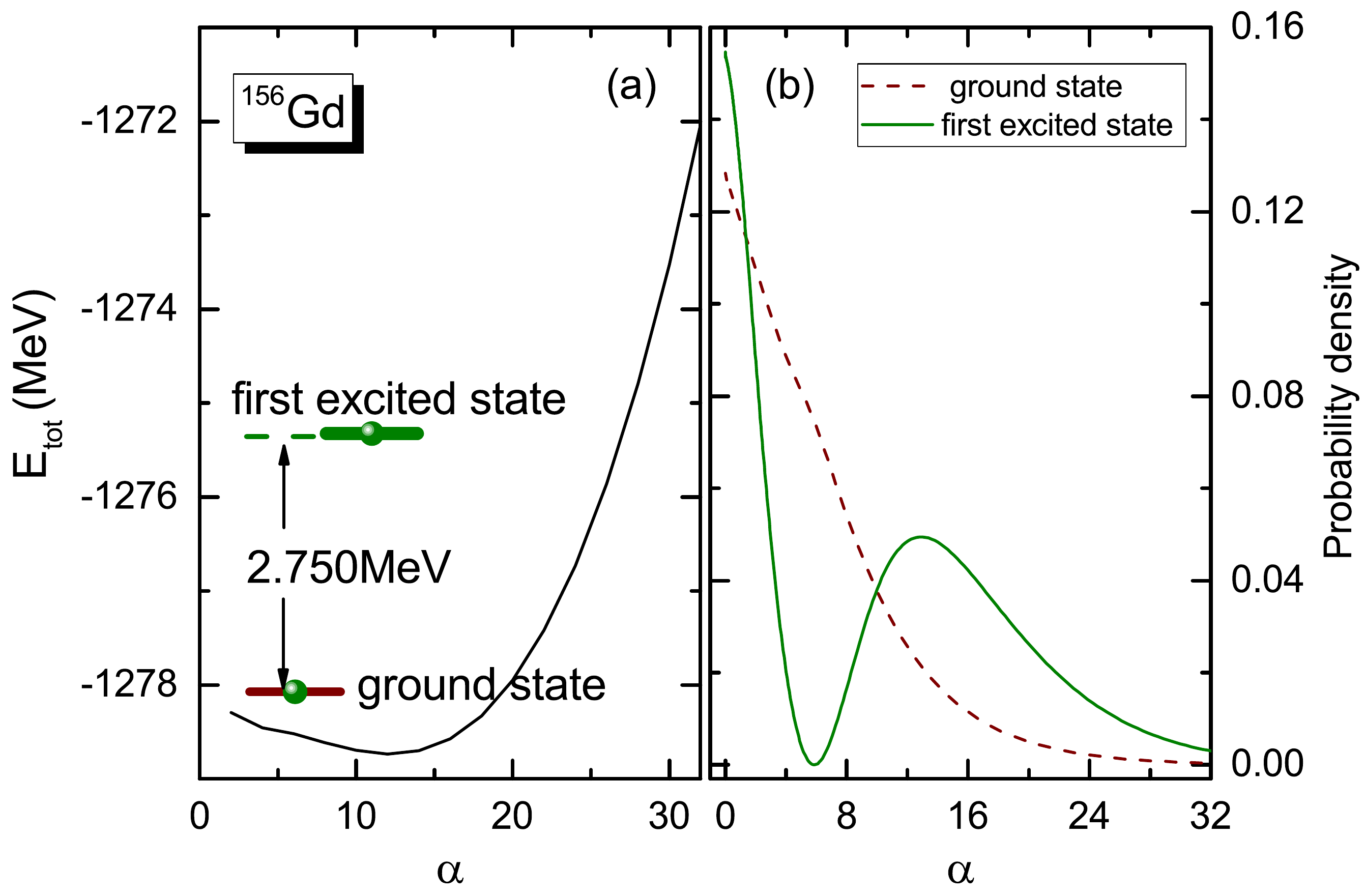}
\caption{\label{Ex-wav-Gd156}(Color online) The lowest two eigenstates of the pairing collective Hamiltonian (a) and the  corresponding probability density distributions (b), for $^{156}$Gd.}
\end{figure}
In Fig. \ref{Ex-wav-Gd156} we display the two lowest eigenstates of the pairing collective Hamiltonian (a) and the corresponding probability density distributions (b)  for $^{156}$Gd. In panel (a) the location of the horizontal lines and positions of the dots indicate the energies and expectation values of the intrinsic pairing deformation $\alpha$, respectively. One notes that the expectation values of $\alpha$ calculated from the collective wave functions of the ground state and first-excited state are 6.09 and 11.01, respectively, considerably smaller than the value of $\alpha$ in the energy minimum. A similar result was also obtained in Ref.~\cite{GozdzNPA1985}. %
%%--------------------------------------------------------------
\subsection{Low-energy structure of $^{156}$Gd calculated with the QPCH}
\begin{figure}[ht]
\centering{\includegraphics[width=0.5\textwidth]{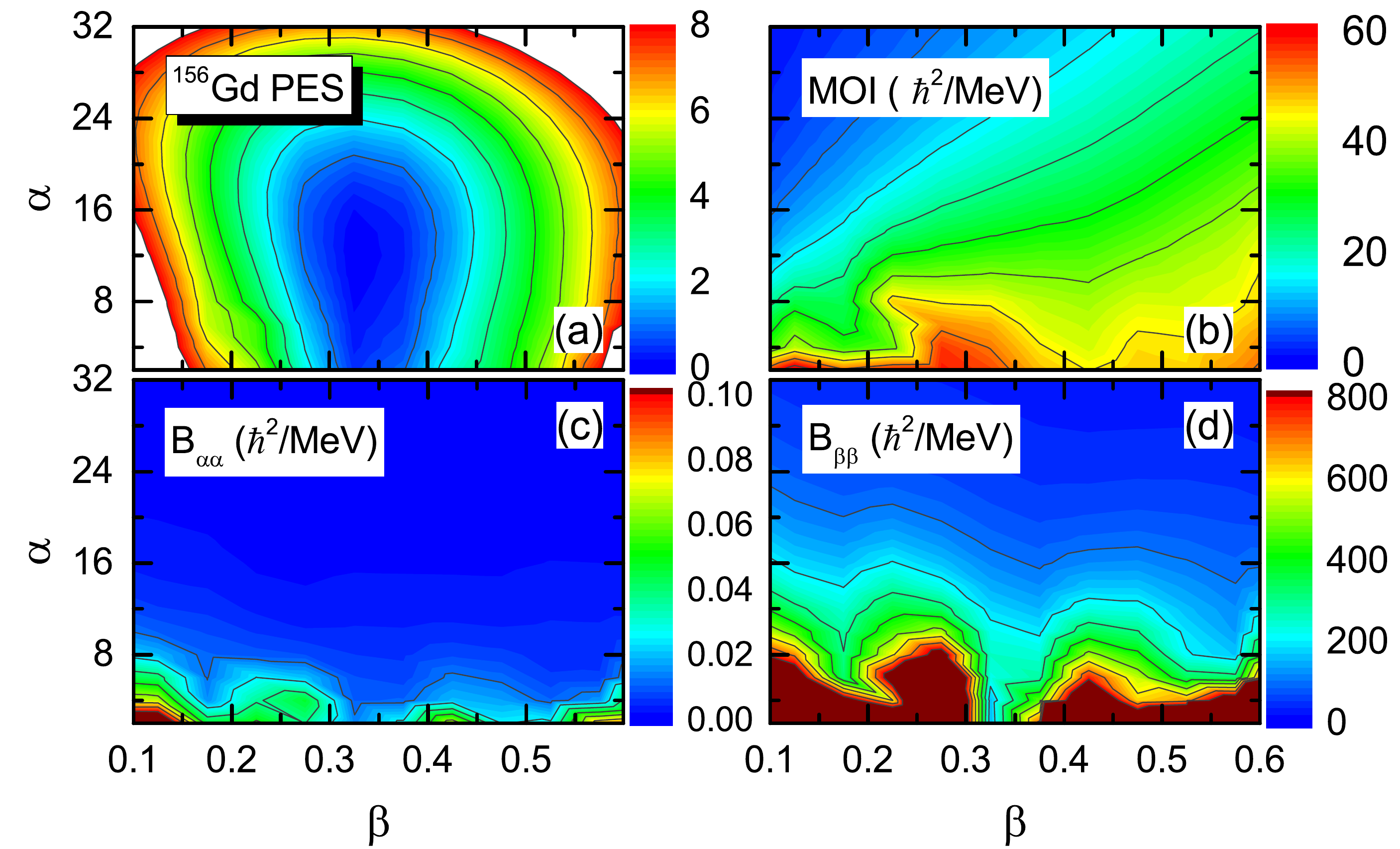}}
\caption{\label{PES-Gd156}(Color online) The potential energy surface (PES), moment of inertia ${\cal I}$, mass parameters $B_{\alpha\alpha}$ and $B_{\beta\beta}$ of $^{156}$Gd in the $(\beta,\alpha)$ plane, calculated by the constrained RMF+BCS with the PC-PK1 energy density functional and separable pairing interaction. All energies (in MeV) in the PES are normalised with respect to the binding energy of the absolute minimum. In all panels the contours join points on the surface with the same values.}
\end{figure}

\begin{figure*}[ht]
\includegraphics[width=0.98\textwidth]{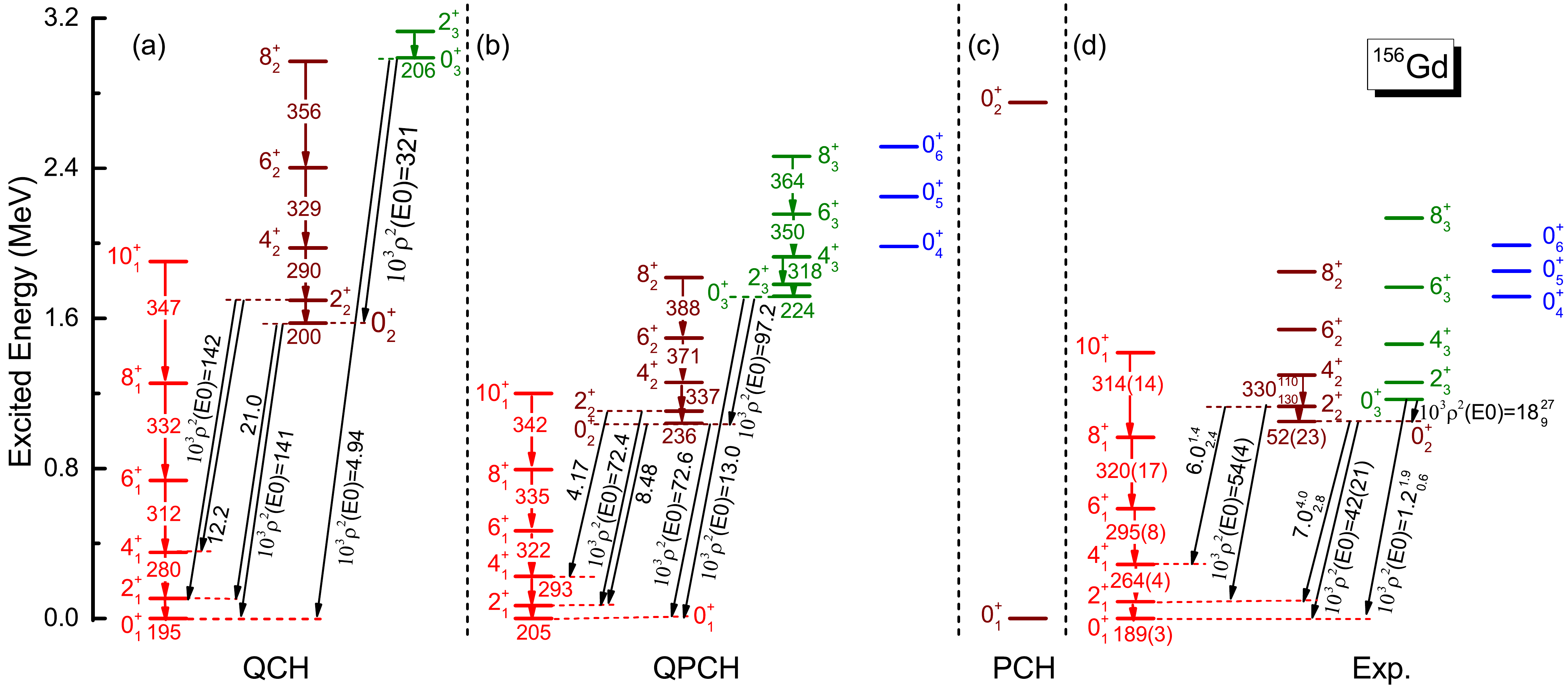}
\caption{\label{Spectra-Gd156}(Color online) The low-energy excitation spectra of $^{156}$Gd, calculated with the QCH (a), QPCH (b), PCH (c), based on the PC-PK1 energy density functional \cite{ZhaoPRC2010}, in comparison with the available data (d) from Refs. \cite{NNDC,AprahamianPRC2018}. The corresponding electric quadrupole and monopole transition probabilities are also compared to data.}
\end{figure*}

Next we consider $^{156}$Gd as a test example for the quadrupole-and-pairing collective Hamiltonian (QPCH) based on relativistic EDFs. The low-energy spectra will also be compared to those obtained with the axially symmetric quadrupole collective Hamiltonian (QCH), which includes only vibrational and rotational dynamic degrees of freedom. Starting from constrained self-consistent RMF+BCS solutions, that is, using the single-particle wave functions, occupation probabilities, and quasiparticle energies that correspond to each point on the energy surface, the parameters that determine the collective Hamiltonian are calculated as functions of quadrupole deformation $\beta$ and pairing deformation $\alpha$. As an illustration, for $^{156}$Gd the potential energy surface (PES), moment of inertia $\cal {I}$ and mass parameters $B_{\alpha\alpha}$ and $B_{\beta\beta}$  are displayed in Fig.~\ref{PES-Gd156}. The global minimum is calculated at $(\beta, \alpha) = (0.325,\ 12)$ and the PES around the minimum appears rather soft, especially with respect to $\alpha$ (cf. also Fig.~\ref{Colpar-Gd156}). The moment of inertia generally increases with the quadrupole deformation $\beta$, while displaying a decrease for larger values of $\alpha$. The mass parameters  $B_{\alpha\alpha}$ and $B_{\beta\beta}$ exhibit a pronounced dependence on $\alpha$, that is, both increase steeply as pairing becomes weaker. This is consistent with the result of Ref.~\cite{PilatNPA1993}.

\begin{figure}[ht]
\includegraphics[width=0.45\textwidth]{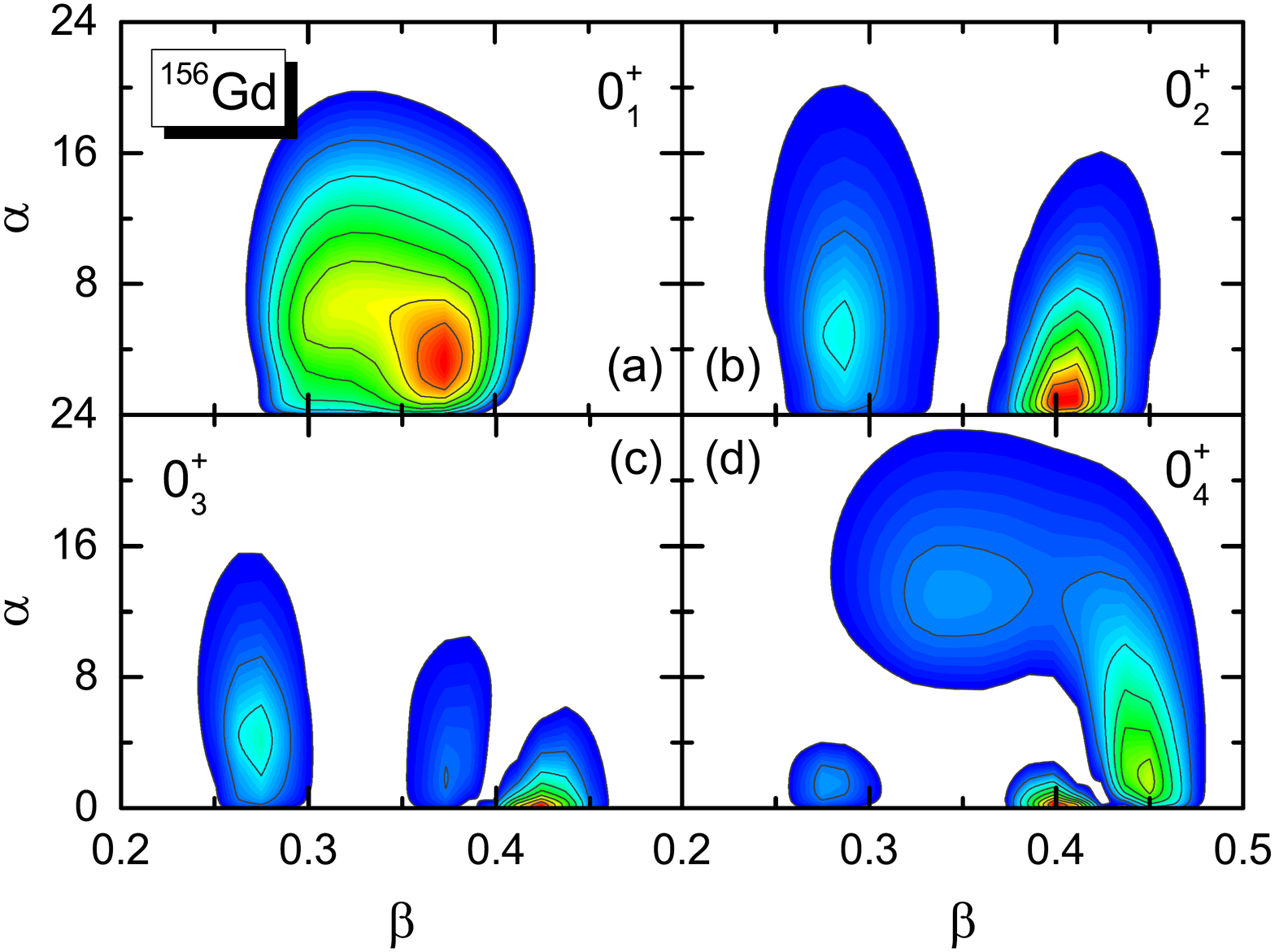}
\caption{\label{Wav-Gd156}(Color online) Probability density distributions in the ($\beta,\alpha$) plane for the first four $0^+$ states of $^{156}$Gd, calculated with the QPCH based on the PC-PK1 energy density functional.}
\end{figure}

The diagonalization of the resulting Hamiltonian yields the excitation energies and collective wave functions for each value of the total angular momentum. In Fig.~\ref{Spectra-Gd156} we plot the QPCH excitation spectrum of $^{156}$Gd, in comparison with available data \cite{NNDC,AprahamianPRC2018}, and results obtained with the QCH and PCH.  In addition to the excitation energies, quadrupole E2 and monopole E0 transition rates are also shown in the figure. Obviously the coupling between shape and pairing dynamical degrees of freedom has a pronounced effect on the calculated spectra. When compared to the results of the QCH model, the inclusion of dynamical pairing increases the moment of inertia and lowers the bands based on excited $0^+$ states, altogether bringing the theoretical spectrum in much better agreement with experiment.
In particular, we note that the $0^+_2$ and $0^+_3$ states are lowered by $\sim0.534$ and $\sim1.271$ MeV, respectively.
The coupling to pairing vibrations increases slightly the intra-band electric quadrupole transition rates, while the calculated E0 rates are generally in better agreement with data. It appears that the QPCH qualitatively reproduces the excitation energies of the first five excited $0^+$ states. The PCH can, of course, only be used to calculate $0^+$ states, and the lowest two have already been shown in Fig.~\ref{Ex-wav-Gd156}. Without coupling to the axial quadrupole deformation, the $0^+_2$ state is predicted at a very high excitation energy.

Figure \ref{Wav-Gd156} displays the probability density distributions in the ($\beta,\alpha$) plane of the first four $0^+$ states of $^{156}$Gd, calculated with the QPCH based on the PC-PK1 energy functional. One finds nodes in the $\beta$ direction for the $0^+_2$, $0^+_3$, and $0^+_4$ states. The distribution of the $0^+_4$ state in the $\alpha$ direction indicates a structure characterized by pairing vibration.

%--------------------------------------------------------------
\subsection{Systematics of low-lying spectra of ${\bf N=92}$ isotones}

\begin{figure}[ht]
\includegraphics[width=0.45\textwidth]{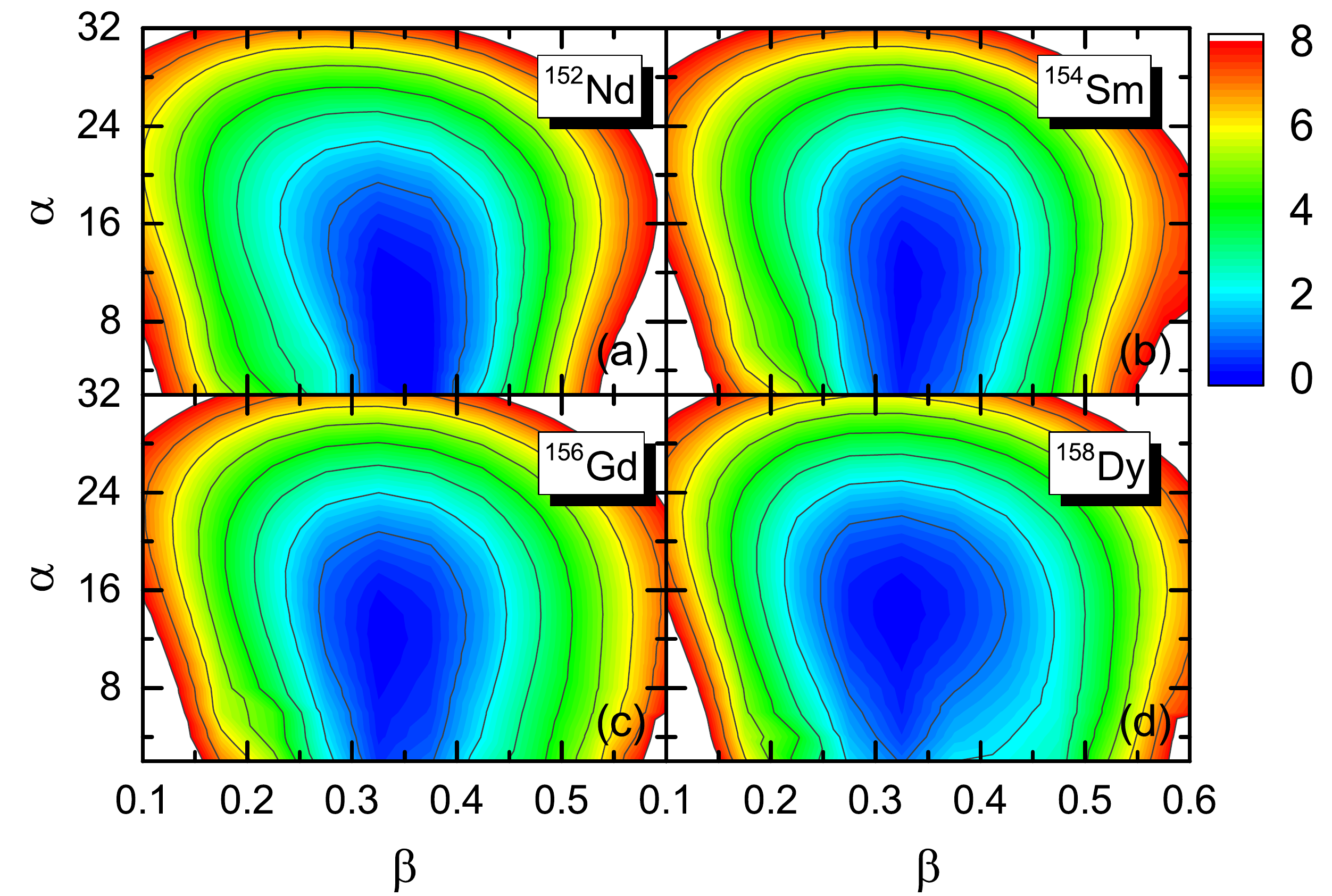}
\caption{\label{PES-N=92}(Color online) The deformation energy surfaces (in MeV) of even-even $N=92$ isotones  in the $(\beta,\alpha)$ plane, calculated using the RMF+BCS model with the PC-PK1 energy functional and separable pairing force.}
\end{figure}

\begin{figure}[ht]
\includegraphics[width=0.45\textwidth]{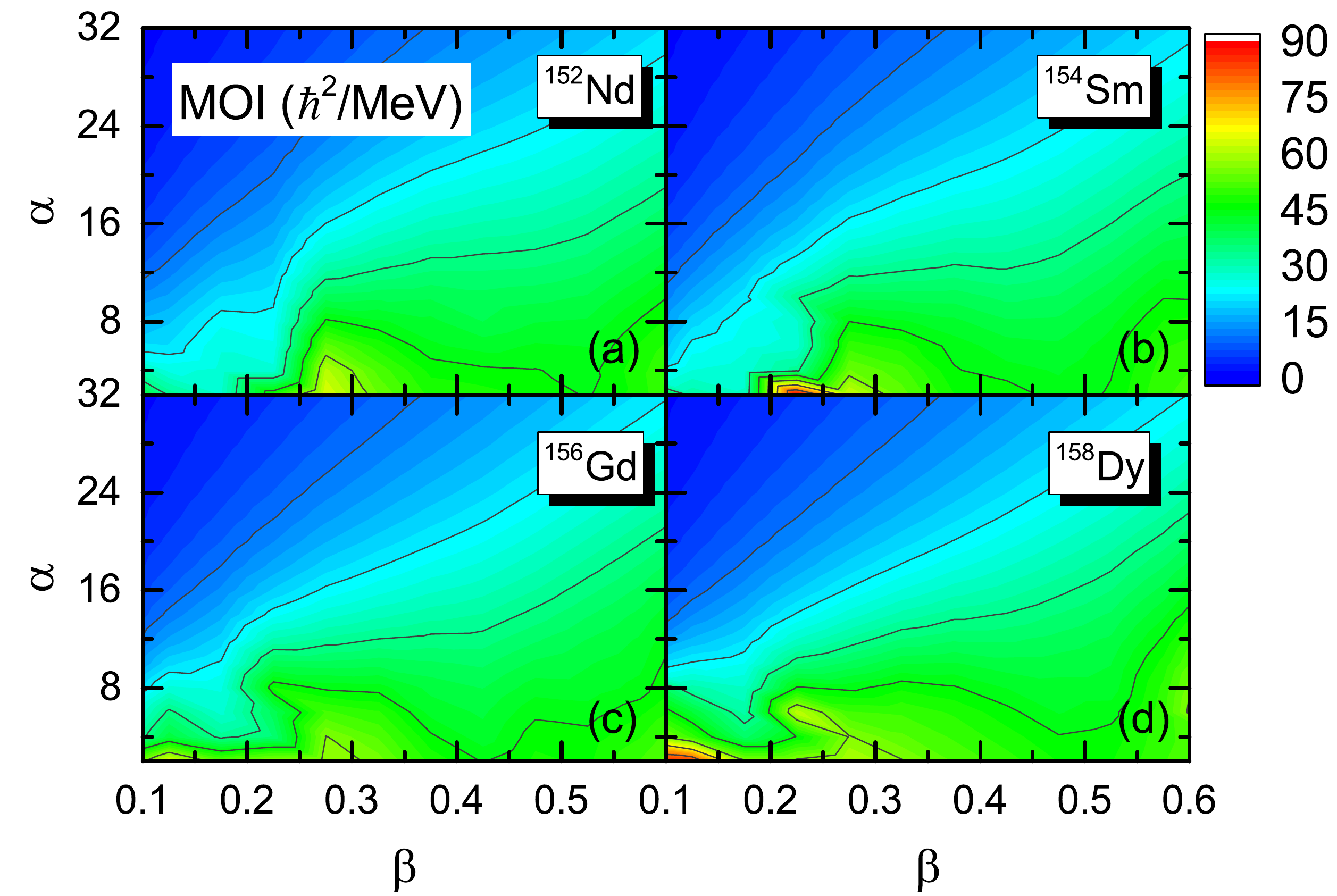}
\caption{\label{Ix-N=92}(Color online) Same as in the caption to Fig. \ref{PES-N=92} but for the moment of inertia.}
\end{figure}

\begin{figure}[ht]
\includegraphics[width=0.45\textwidth]{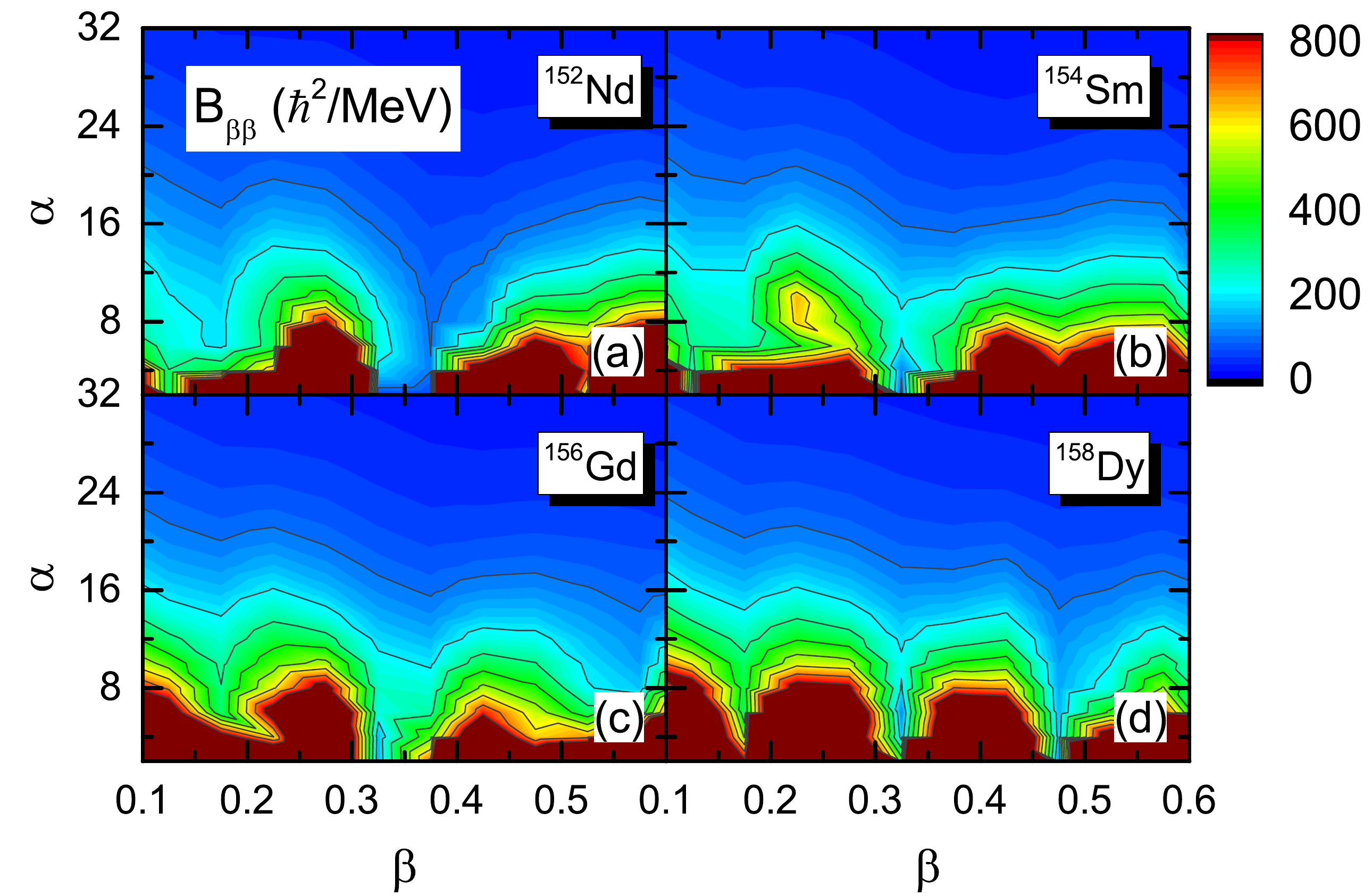}
\caption{\label{Bbb-N=92}(Color online) Same as in the caption to Fig. \ref{PES-N=92} but for the collective mass $B_{\beta\beta}$.}
\end{figure}

\begin{figure}[ht]
\includegraphics[width=0.45\textwidth]{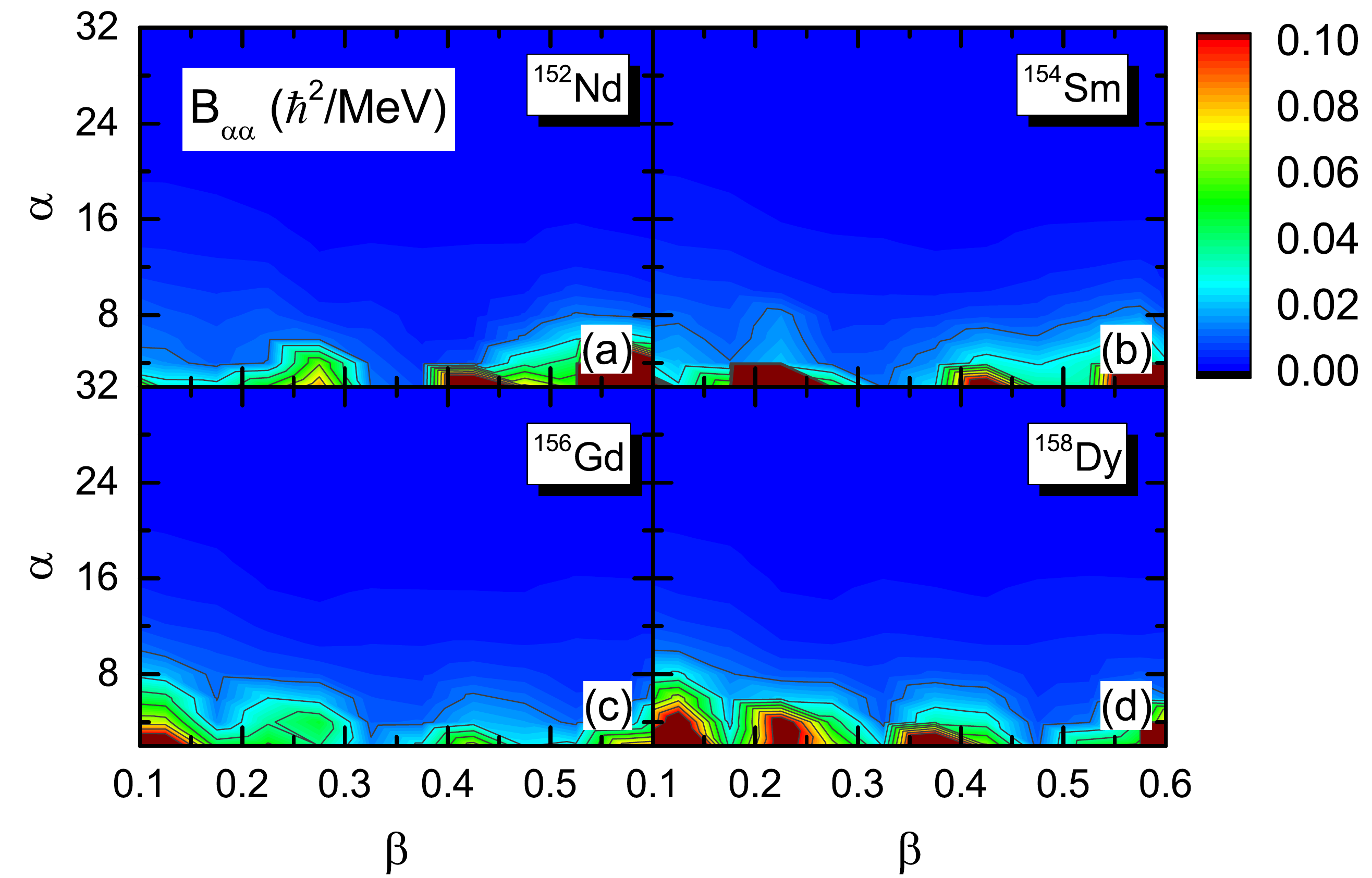}
\caption{\label{Baa-N=92}(Color online) Same as in the caption to Fig. \ref{PES-N=92} but for the collective mass $B_{\alpha\alpha}$.}
\end{figure}

As a further test of the QPCH we analyze the systematics of the low-lying spectra of four even-even axially deformed $N=92$ isotones: $^{152}$Nd, $^{154}$Sm,
$^{156}$Gd and $^{158}$Dy.
Figure \ref{PES-N=92} displays the deformation energy surfaces of the $N=92$ isotones, calculated with the PC-PK1 energy density functional and separable pairing force. The energy surfaces exhibit pronounced global minima for a rather large value of the quadrupole deformation $\beta \approx 0.35$, and pairing deformation $\alpha \approx  12 - 16$. The minima appear quite soft  towards smaller values of the pairing collective coordinate $\alpha$. It is interesting to note that this softness is reduced with the increase of the proton number, while simultaneously the energy surfaces become more soft in the quadrupole collective deformation.
The moment of inertia $\cal{I}$, and the collective masses $B_{\beta\beta}$ and $B_{\alpha\alpha}$, are displayed in Figs. \ref{Ix-N=92}-\ref{Baa-N=92}, respectively. The collective parameters of the isotones $^{152}$Nd, $^{154}$Sm, and $^{158}$Dy present patterns very similar to those of $^{156}$Gd, already discussed in the previous section.

\begin{figure}[ht]
\includegraphics[width=0.45\textwidth]{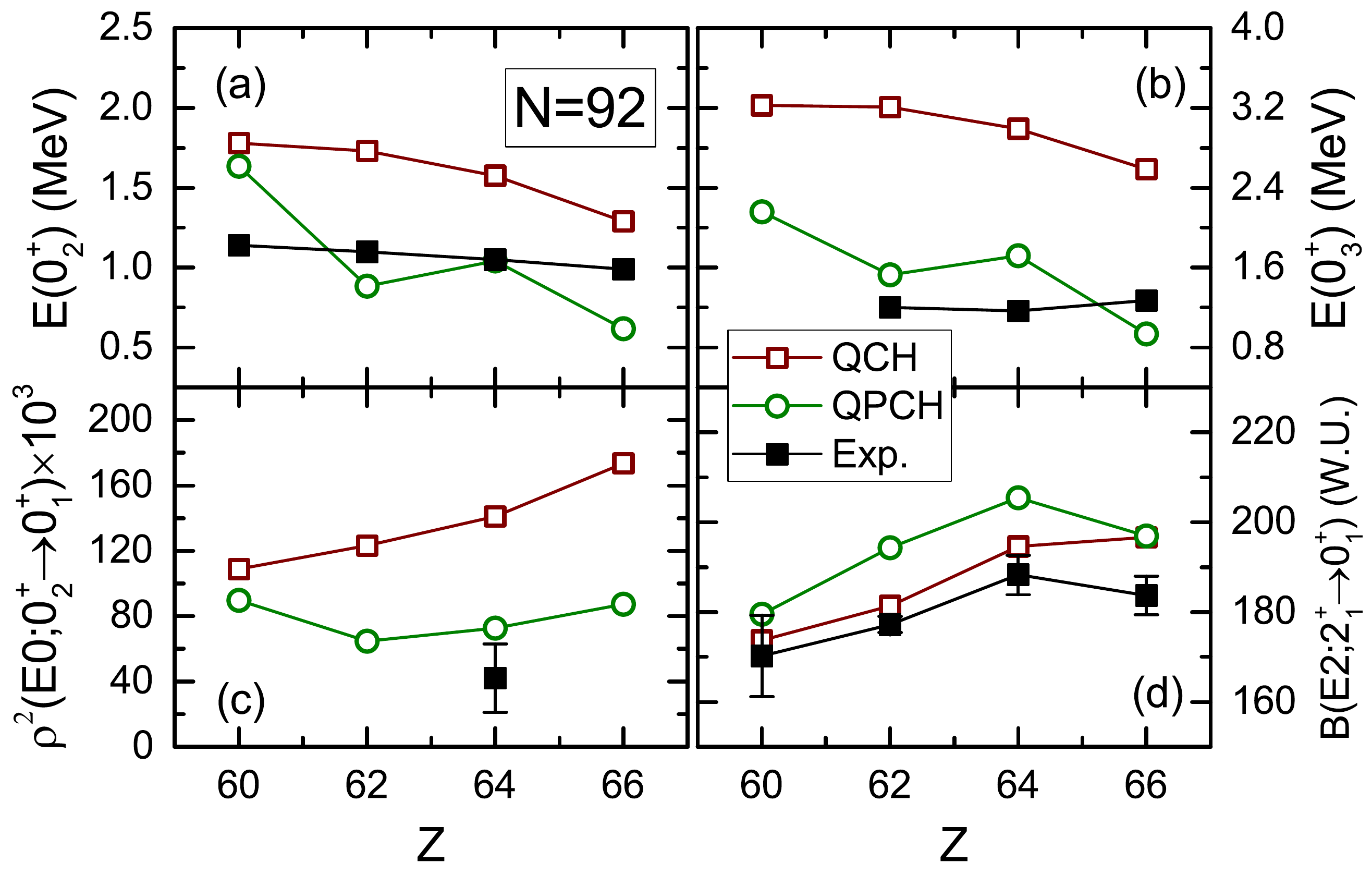}
\caption{\label{Obs-N=92}(Color online) The excitation energies of the two lowest excited $0^+_2$ (a) and $0^+_3$ (b) states, the E0-transition strengths $\rho^2(E0;0^+_2\rightarrow0^+_1)$ (c), and the $B(E2;2^+_1\rightarrow0^+_1)$ values (d) of four even-even $N=92$ isotones, as functions of the proton number. Values calculated with the QPCH based on the PC-PK1 functional are shown in comparison with the available data \cite{NNDC,KibediADNDT2005}, and corresponding results obtained with the usual QCH.}
\end{figure}

The effect of pairing vibrations are further illustrated in the systematics of low-lying spectra of $N=92$ isotones. Figure  \ref{Obs-N=92} displays the evolution of excitation energies of the two lowest excited $0^+_2$ and $0^+_3$ states, the E0-transition strengths $\rho^2(E0;0^+_2\rightarrow0^+_1)$, and the $B(E2;2^+_1\rightarrow0^+_1)$ values of the four even-even $N=92$ isotones, as functions of the proton number. Based on the PC-PK1 functional and separable pairing force, values predicted by the QPCH are compared to those obtained with the usual QCH, that only includes the axial quadrupole deformation as collective coordinate, and with available data \cite{NNDC,KibediADNDT2005}. Except for $0^+_2$ in $^{152}$Nd, all the $0^+_2$ and $0^+_3$ states are significantly lowered by the coupling to pairing vibrations, in very good agreement with the experimental excitation energies. The lowering of the $0^+_2$ level ranges from $\sim 0.1$ to $\sim 0.9$ MeV in the present calculation, while for the $0^+_3$ state the interval is $\sim1.2$ to $\sim1.7$ MeV.

The E0 transition probabilities from the $0^+_2$ state to the ground state are considerably reduced by the inclusion of the dynamical pairing degree of freedom, while the QPCH calculation moderately increases the  $B(E2;2^+_1\rightarrow0^+_1)$ values of $^{152}$Nd, $^{154}$Sm and $^{156}$Gd, with respect to the B(E2)'s predicted by the QCH model. However, all calculated transition rates are generally in good agreement with data, especially considering that only one shape deformation degree of freedom is taken into account.

\begin{figure}[ht]
\includegraphics[width=0.35\textwidth]{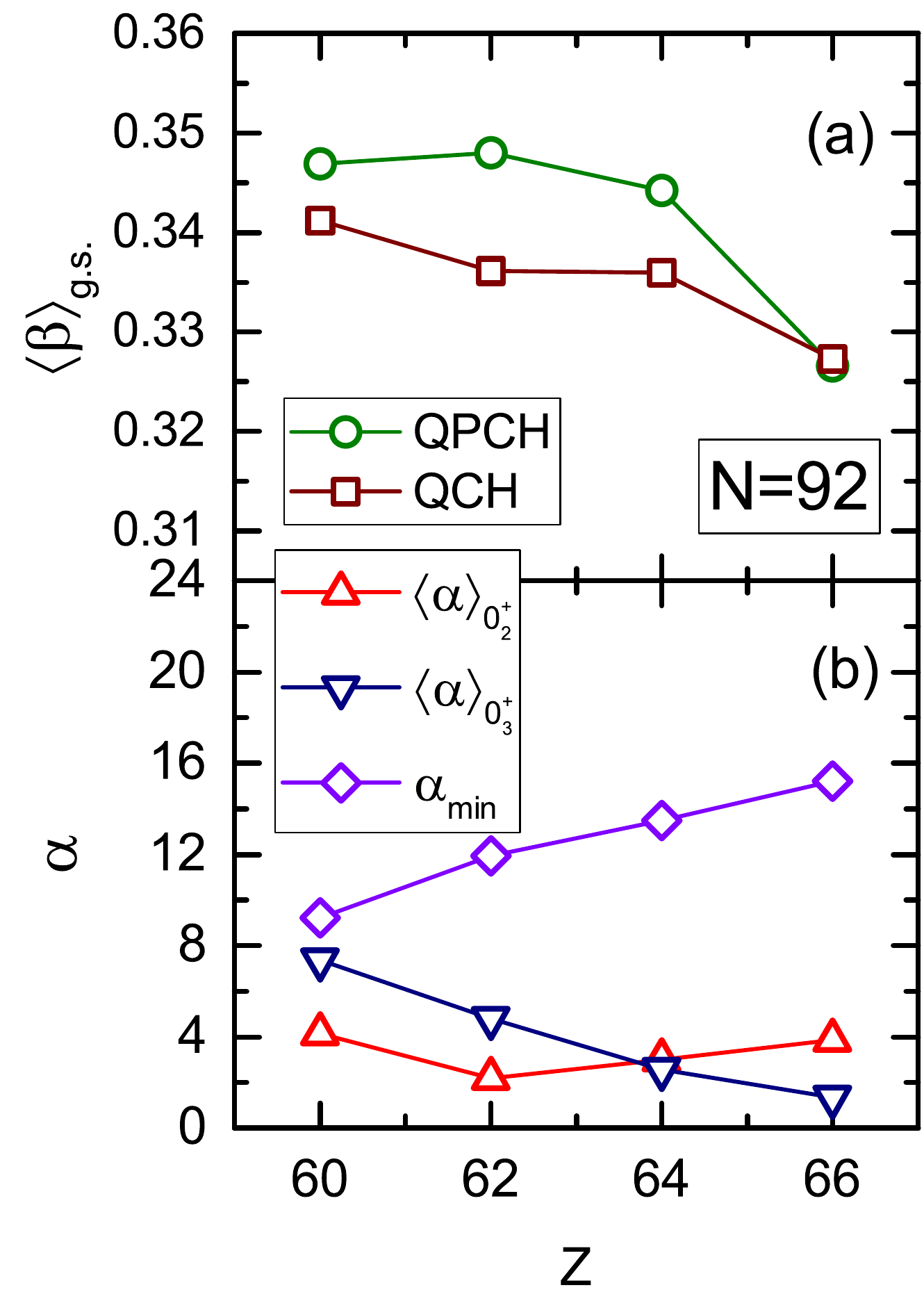}
\caption{\label{ObsExpr-N=92}(Color online) (a) The expectation value of the quadruple deformation $\langle\beta\rangle$ for the ground states of four $N=92$ isotones, calculated with the QPCH and QCH, respectively. (b) The self-consistent minima $\alpha_{\rm min}$ of the potential energy surfaces, and the expectation value of the pairing deformation $\langle\alpha\rangle$ for the $0^+_2$ and $0^+_3$ states of the four $N=92$ isotones.}
\end{figure}

It is interesting to consider these results in relation to the trend exhibited by the collective coordinates. In Fig. \ref{ObsExpr-N=92} (a), we plot the expectation values of the quadrupole deformation $\langle\beta\rangle$ for the ground states of the four $N=92$ isotones calculated with QPCH and QCH, respectively. The expectation values of the pairing deformation $\langle\alpha\rangle$ calculated with the QPCH for the $0^+_2$ and $0^+_3$ states are  also compared with the self-consistent minima $\alpha_{\rm min}$ of the PESs shown in Fig. \ref{ObsExpr-N=92} (b). One notes that $\langle\beta\rangle$ increases slightly for $^{152}$Nd, $^{154}$Sm, and $^{156}$Gd when dynamical pairing is included. This is consistent with the variation of $B(E2; 2^+_1\to 0^+_1)$ in these nuclei. As already noted, the expectation values $\langle\alpha\rangle$ for $0^+_2$ and $0^+_3$ are considerably smaller than the equilibrium value and, consequently, this leads to a significant increase of inertia masses for these states and lowers the corresponding excitation energies. 

%\red{It is interesting to consider these results in relation to the trend of collective coordinates. In Fig. \ref{ObsExpr-N=92}(a) we plot The expectation value of quadruple deformation $\beta$ for $0^+_1$ states of the four $N=92$ isotones, calculated by QPCH and QCH, respectively. The self-consistent minima $\alpha_{min}$ of the potential energy surfaces, and the expectation value of pairing deformation $\alpha$ for $0^+_1$ states of the four $N=92$ isotones are also displayed in Fig.~\ref{ObsExpr-N=92}(b). As already noted, the expectation value $\langle\alpha\rangle$ in the ground state is considerably smaller than the equilibrium value and, consequently, this leads to a significant increase of $B_{\alpha\alpha}$ for the ground-state. The effect is the lowering of excitation energies of the  $0^+$ states.}
%

\section{\label{Sec:IIII} Summary and Outlook}
The quadrupole collective Hamiltonian (QCH), based on the framework of microscopic energy density functionals, has been extended to include a pairing collective coordinate. In addition to quadrupole shape vibrations and rotations, the model describes pairing vibrations and explicitly couples shape and pairing degrees of freedom. The parameters of the kinetic term of the quadrupole-and-pairing collective Hamiltonian (QPCH), as well as the potential, are determined by fully self-consistent mean-field calculations, with constraints on the collective coordinates.

In the particular implementation considered in the present work, we have simplified the model by assuming axial shape symmetry and neglecting pairing rotations. Therefore, only two intrinsic collective coordinates have explicitly been considered: the axial quadrupole deformation $\beta$, and the intrinsic pairing deformation $\alpha$ related to the gap parameter $\Delta$.
Without such simplifications, that is, by including additional collective coordinates, the method quickly becomes computationally excessive, especially for heavy nuclei. As our aim here is to test the effect of coupling between shape and pairing degrees of freedom on low-energy spectra, the specific model includes only two intrinsic collective coordinates.

Constrained self-consistent mean-field calculations in the $(\beta,\alpha)$ plane have been performed for four $N=92$ axially deformed rare-earth isotones, using the PC-PK1 relativistic density functional in the particle-hole channel, and pairing correlations are included in the BCS approximation with a pairing force separable in momentum space. The resulting single-nucleon wave functions, energies and occupation probabilities, as functions of the intrinsic deformations $\beta$ and $\alpha$, provide the microscopic input for the parameters of the collective Hamiltonian: three mass parameters $B_{\beta\beta}$, $B_{\alpha\alpha}$, and $B_{\beta\alpha}$, the moment of inertia $\cal{I}$, and the collective potential. The moments of inertia are calculated using the Inglis-Belyaev formula, and the mass parameters associated with the collective coordinates $\beta$ and $\alpha$ are computed in the cranking approximation. An extensive test has been carried out in calculations of potential energy surfaces, and the resulting collective excitation spectra and transition probabilities. Results for excitation energies in the ground-state band and bands based on excited $0^+$ states, the corresponding intra-band and inter-band E2 transition probabilities, as well as E0 transition rates, have been compared to available data and values obtained using the standard quadrupole collective Hamiltonian (QCH). The effect of the inclusion of pairing vibrations on low-lying excitation spectra has been analyzed for the four $N=92$ isotones: $^{152}$Nd, $^{154}$Sm,
$^{156}$Gd and $^{158}$Dy.

The analysis has demonstrated, in a quantitative way, the importance of the dynamical pairing degree of freedom. Even though in several studies of low-energy collective spectra this effect was taken into account in an approximative way,
here we have explicitly considered the coupling between shape and pairing degrees of freedom in the parameters that determine the collective Hamiltonian. It has been shown that the coupling to pairing vibrations increases the moment of inertia, lowers the energies of excited $0^+$ states and bands built on them, reduces the E0-transition strengths and, generally, produces low-energy spectra in much better agreement with experimental results.

The present study has been restricted to axially symmetric nuclei, and we have only analyzed the low-energy spectra of four rare-earth nuclei. As the effect of pairing vibrations will particularly be important for nuclei characterized by shape coexistence, it is essential to extend the current implementation of the model to include the triaxial degree of freedom. Future applications will consider other mass regions and, in particular, soft nuclei that exhibit quantum shape-phase transitions. Another interesting development will be the extension of the model to heavy nuclei characterized by pronounced octupole correlations.

\begin{acknowledgments}
This work has been supported in part by the NSFC under Grants No. 11765015, No. 11875225,  No. 11675065, and No. 11790325, Joint Fund Project of Education Department in Guizhou Province(No. Qian Jiao He KY Zi[2016]312, Qianjiaohe KY Zi[2018]433), Qiannan normal University Initial Research Foundation Grant to Doctor(qnsyrc201617), the science and technology program foundation of Guizhou province(Qian KeHe Platform Talents[2019]QNSYXM-03), and by the QuantiXLie Centre of Excellence, a project co-financed by the Croatian Government and European Union through the European Regional Development Fund - the Competitiveness and Cohesion Operational Programme (KK.01.1.1.01).
\end{acknowledgments}

\appendix

\section{Diagonalization of the collective Hamiltonian}\label{Sec:ADCH}
\begin{figure}[h]
\centering{\includegraphics[width=0.45\textwidth]{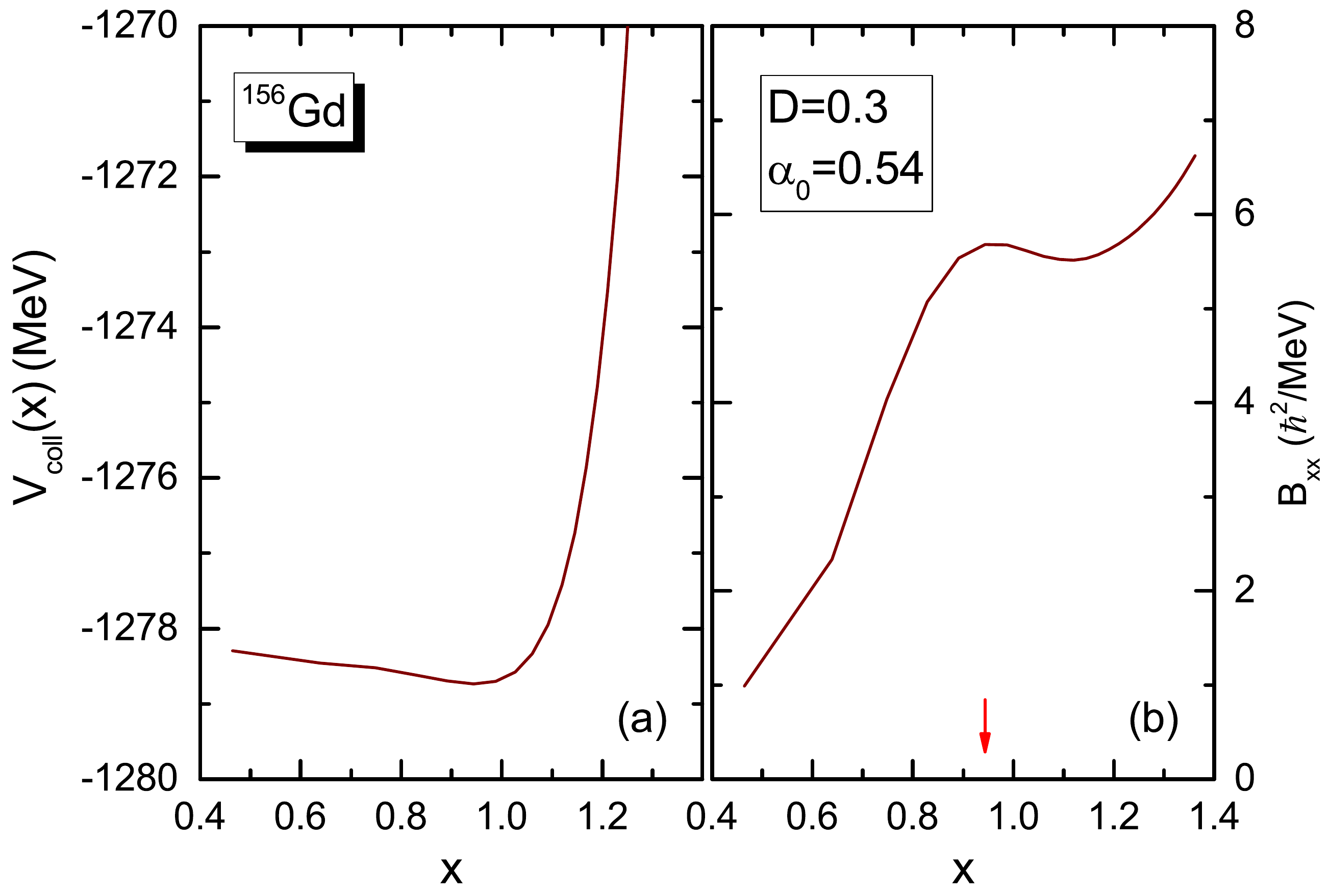}}
\caption{\label{VxBxx-Gd156}(Color online) The energy (a) and collective mass $B_{xx}$(b) of $^{156}$Gd, as functions of $x$. The location of the energy minimum is indicated with the red arrow.}
\end{figure}

\begin{figure}[h]
\includegraphics[width=0.45\textwidth]{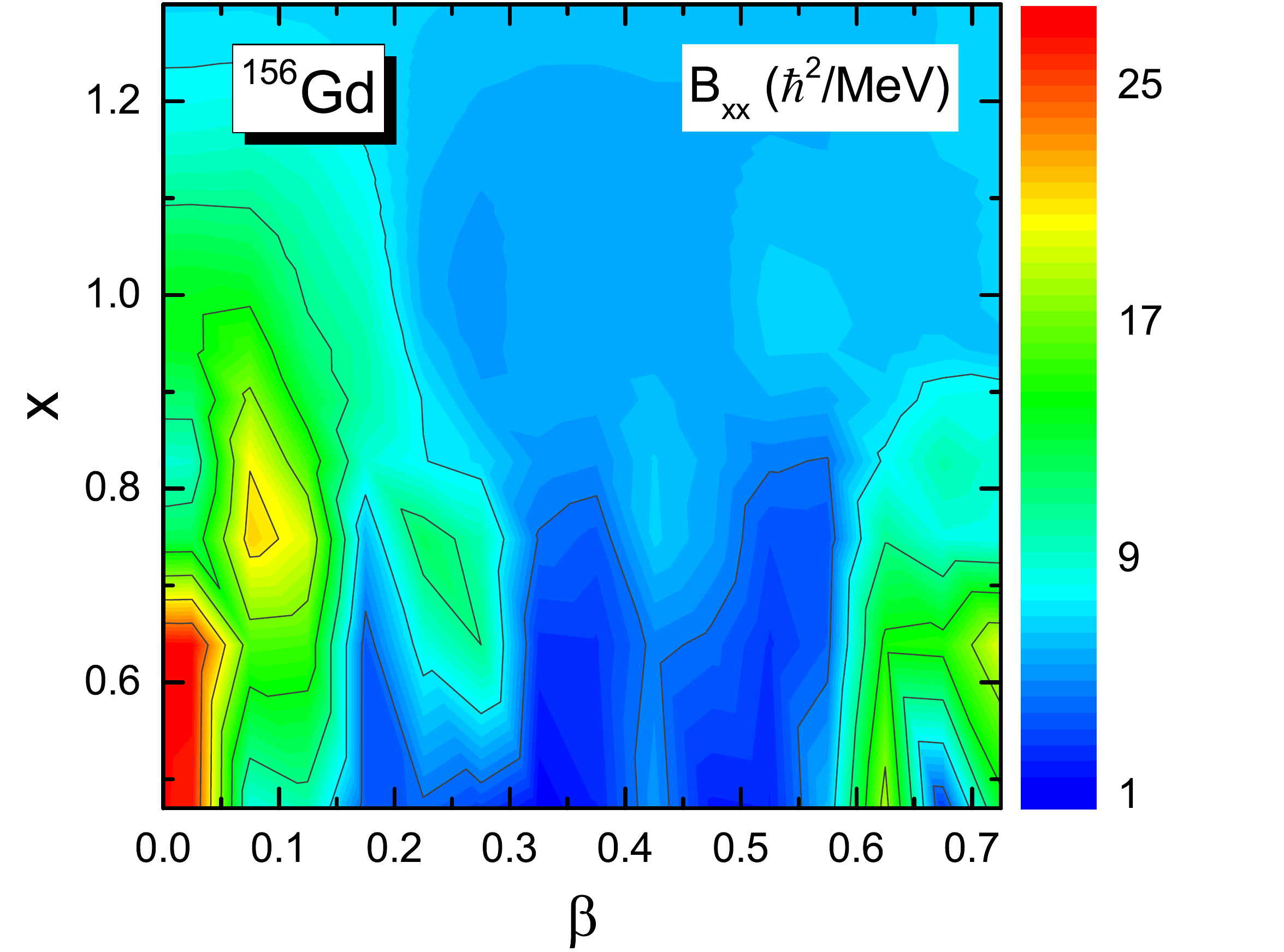}
\caption{\label{Bxx-Gd156}(Color online) The collective mass $B_{xx}$ in the $(\beta,x)$ plane of $^{156}$Gd.}
\end{figure}

In Figs. \ref{Colpar-Gd156} and \ref{PES-Gd156} it has been shown that  the collective mass $B_{\alpha\alpha}$ exhibits a very pronounced dependence on the pairing deformation $\alpha$. $B_{\alpha\alpha}$ decreases very steeply with the increase of $\alpha$ and, because the collective mass appears in the denominator of the kinetic term of the collective Hamiltonian Eq.~(\ref{eq:PCH}), this leads to a slow convergence when the eigenfunctions are expanded in the harmonic-oscillator basis \cite{GozdzNPA1985}. The diagonalization of the collective Hamiltonian is here illustrated with the example of the pairing Hamiltonian PCH. As shown in Ref.~\cite{GozdzNPA1985}, a much better convergence can be reached when one performs a transformation from $\alpha$ to a new coordinate $x$
\begin{equation}
 x=D\ln\left(1+\frac{\alpha}{\alpha_0}\right),
 \end{equation}
 and the collective mass $B_{\alpha\alpha}$ is then transformed to $B_{xx}$
 \begin{equation}\label{eq:Bxx}
 B_{xx}=B_{\alpha\alpha}\frac{(\alpha+\alpha_0)^2}{D^2}.
 \end{equation}
The parameters $D$ and $\alpha_0$ are determined by fitting the collective mass $B_{\alpha\alpha}$ to $M_{\alpha\alpha}$
\begin{equation}
M_{\alpha\alpha}\approx\frac{D^2}{(\alpha+\alpha_0)}.
\end{equation}
Fig.~\ref{VxBxx-Gd156} displays the collective potential $V_{\rm coll}(x)$ and $B_{xx}$ of $^{156}$Gd, as functions of the new coordinate x. For the two-dimensional (2D) case which includes the axial quadrupole deformation, in Fig.~\ref{Bxx-Gd156} we plot the collective mass $B_{xx}$ in $(\beta,x)$ plane of $^{156}$Gd. The variation of $B_{xx}$ with the coordinate x is more smooth than that of $B_{\alpha\alpha}$ (cf. Fig. \ref{PES-Gd156}).

Expressed in the new coordinate the pairing collective Hamiltonian takes the form
\begin{equation}\label{eq:PCH-x}
{\hat H}_{\rm coll}(x)=-\frac{\hbar^2}{2}\frac{1}{\sqrt{B_{xx}}}\frac{\partial}{\partial x}\frac{1}{\sqrt{B_{xx}}}\frac{\partial}{\partial x} + V_{\rm coll}(x)\;,
\end{equation}
and thus the basis states used to diagonalize the Hamiltonian (\ref{eq:PCH-x}), will be generated by the harmonic oscillator Hamiltonian
\begin{equation}\label{eq:BaseH}
\hat{H}_B=-\frac{\hbar^2}{2\mu}\frac{d^2}{dx^2} + \frac{1}{2}\mu\omega^2x^2\;.
\end{equation}
The corresponding oscillator length parameter reads
\begin{equation}
b_x=\sqrt{\frac{\hbar}{M_x\omega_x}}\;.
\end{equation}
The eigenfunctions of the Hamiltonian (\ref{eq:BaseH})
\begin{equation}\label{basisx}
\phi_{n_x}(x) = \frac{N_{n_x}}{b_x}H_{n_x}\left(\zeta\right)e^{-\zeta^2/2}\;,
\end{equation}
correspond to the Hermite polynomials $H_{n_x}(\zeta)$, where $ \zeta=x/b_x$. In the two-dimensional case a harmonic oscillator basis is also used for the expansion of the eigenfunctions of the quadrupole shape collective coordinate.

For the self-consistent mean-field calculation the Dirac equation is solved by expanding the spinors in terms of a harmonic oscillator basis with 14 major shells. The RMF+BCS equations are solved on a mesh in the $\beta-\alpha$ plane
\begin{equation}\label{eq:mesh}
-0.025\leq\beta\leq0.775,\ \ 2\leq\alpha\leq 50
\end{equation}
with steps 0.05 and 2, respectively.

In the calculation of the matrix elements of the collective Hamiltonian, with the substitution $y\equiv\beta b_{\beta}$,  the integrals over $\beta$ are evaluated by Gauss-Laguerre quadrature. The integrals over $\alpha$ are evaluated by Gauss-Legendre quadrature, with  the substitution $z\equiv xb_x$. The corresponding number of mesh points are $n_\beta=64$ and $n_x=520$, respectively. The parameters of the collective Hamiltonian at the Gaussian mesh points are determined by interpolation from the values calculated on the equidistant mesh defined by Eq. (\ref{eq:mesh}).

\bibliography{reference}% Produces the bibliography via BibTeX.

\end{document}